\documentclass[onecolumn,showpacs,nofootinbib,preprintnumbers,superscriptaddress]{revtex4}

\usepackage{epsfig}
\usepackage{bm}
\usepackage{amsfonts}
\usepackage{amssymb,amsmath}
\usepackage{enumerate}
\usepackage{multirow}
\usepackage{epsfig}
\usepackage{graphicx}
\usepackage{bm}

\def\be{\begin{equation}}
\def\ee{\end{equation}}
\def\bea{\begin{eqnarray}}
\def\eea{\end{eqnarray}}

\renewcommand{\(}{\left(}
\renewcommand{\)}{\right)}
\renewcommand{\[}{\left[}
\renewcommand{\]}{\right]}

\begin{document}

\title{Observational constraints on phantom power-law cosmology}

\author{Chakkrit Kaeonikhom} \email{kchakkrit@nu.in.th}
\affiliation{Department of Physics, Naresuan University, Phitsanulok 65000,
Thailand}
\affiliation{Institute for Fundamental Study (TPTP-IF), Naresuan University, Phitsanulok 65000,
Thailand}

\author{Burin Gumjudpai\footnote{Corresponding author}}
\email{buring@nu.ac.th}\affiliation{Department of Physics, Naresuan University, Phitsanulok 65000,
Thailand}\affiliation{Institute for Fundamental Study (TPTP-IF), Naresuan University, Phitsanulok 65000,
Thailand}\affiliation{Thailand Center of Excellence in Physics\\CHE, Ministry of Education, Bangkok 10400, Thailand}
 \affiliation{National Astronomical Research Institute of Thailand\\Siripanich Bld., 191 Huay Kaew Rd.,  Chiang Mai 50200, Thailand}

\author{Emmanuel N. Saridakis}
\email{msaridak@phys.uoa.gr}
 \affiliation{College of Mathematics
and Physics\\Chongqing University of Posts and
Telecommunications, Chongqing 400065, P.R. China}

\begin{abstract}
We investigate phantom cosmology in which the scale factor is a
power law, and we use cosmological observations from Cosmic
Microwave Background (CMB), Baryon Acoustic Oscillations (BAO) and
observational Hubble data, in order to impose complete constraints
on the model parameters. We find that the power-law exponent is
$\beta\approx-6.51^{+0.24}_{-0.25}$, while the Big Rip is realized
at $t_s\approx104.5^{+1.9}_{-2.0}$ Gyr, in 1$\sigma$ confidence
level. Providing late-time asymptotic expressions, we find that
the dark-energy equation-of-state parameter at the Big Rip remains
finite and equal to $w_{DE}\approx -1.153$, with the dark-energy
density and pressure diverging. Finally, we reconstruct the
phantom potential.
\end{abstract}

\pacs{98.80.-k,95.36.+x}

 \maketitle

\section{Introduction}

Recent cosmological observations obtained by SNIa {\cite{c1}},
WMAP {\cite{c2}}, SDSS {\cite{c3}} and X-ray {\cite{c4}} indicate
that the observable universe experiences an accelerated expansion.
Although the simplest way to explain this behavior is the
consideration of a cosmological constant \cite{c7}, the known
fine-tuning problem \cite{8} led to the dark energy paradigm. The
dynamical nature of dark energy, at least in an effective level,
can originate from a variable cosmological ``constant''
\cite{varcc}, or from various fields, such is a canonical scalar
field (quintessence) \cite{quint}, a phantom field, that is a
scalar field with a negative sign of the kinetic term
\cite{Caldwell:2003vq,phant}, or the combination of quintessence
and phantom in a unified model named quintom \cite{quintom}.
Finally, an interesting attempt to probe the nature of dark energy
according to some basic quantum gravitational principles is the
holographic dark energy paradigm \cite{holoext} (although the
recent developments in Horava gravity could offer a dark energy
candidate with perhaps better quantum gravitational foundations
\cite{Horawa}).

The advantage of phantom cosmology, either in its simple or in its
quintom extension, is that it can describe the phantom state of
the universe, that is when the dark energy equation-of-state
parameter lies below the phantom divide $-1$, as it might be the
case according to observations \cite{c1}. Additionally, a usual
consequence of phantom cosmology in its basic form, is the future
Big Rip \cite{BigRip} or similar singularities \cite{BigRip2}, and
thus one needs additional non-conventional mechanism if he desires
to avoid such  a possibility \cite{Sami:2003xv}.

On the other hand power-law cosmology, where the scale factor is a
power of the cosmological time, proves to be a very good
phenomenological description of the universe evolution, since
according to the value of the exponent it can describe the
radiation epoch, the dark matter epoch, and the accelerating, dark
energy epoch \cite{Kolb1989,Peebles:1994xt,Nojiri:2005pu}.
Although it is tightly constrained by nucleosynthesis
\cite{Sethi:1999sq, Kaplinghat:1999}, considering  late universe, it is found to be consistent with age of
high-redshift objects such as globular clusters
\cite{Lohiya:1997ti}, with the SNIa data
\cite{Sethi2005,Dev:2008ey}, and with X-ray gas mass fraction
measurements of galaxy clusters \cite{Allen2002, Zhu:2007tm}.
Furthermore, in the context of the power-law model, one can
describe the gravitational lensing statistics \cite{Dev:2002sz},
the angular size-redshift data of compact radio sources
\cite{Alcaniz:2005ki}, and the SNIa magnitude-redshift relation
\cite{Dev:2002sz, Sethi2005}.

In this work we desire to impose observational constraints on
phantom power-law cosmology, that is on the scenario of a phantom
scalar field along with the matter fluid in which the scale factor
is a power law. In particular, we use cosmological observations
from Cosmic Microwave Background (CMB), Baryon Acoustic
Oscillations (BAO) and observational Hubble data ($H_0$), in order
to impose complete constraints on the model parameters, focusing
on the power-law exponent and on the Big Rip time.

This paper is organized as follows. In section \ref{model} we
construct the scenario of phantom power-law cosmology. In section
\ref{Observational constraints} we use observational data in order
to impose constraints on the model parameters, and in section
\ref{results} we discuss the physical implications of the obtained
results. Finally, section \ref{Conclusions} is devoted to the
conclusions.\\

\section{Phantom cosmology with Power-Law Expansion}
\label{model}

In this section we present phantom cosmology under power-law
expansion. Throughout the work we consider the homogenous and
isotropic Friedmann-Robertson-Walker (FRW) background geometry
with metric
\begin{equation}
ds^2=dt^2-a^2(t)\left[\frac{dr^2}{1-kr^2}+r^2d\Omega_2^2\right],
\end{equation}
where $t$ is the cosmic time, $r$ is the spatial radius
coordinate, $\Omega_2$ is the 2-dimensional unit sphere volume,
and $k$ characterizes the curvature of 3-dimensional space of
which $k=-1,0,1$ corresponds to open, flat and closed universe
respectively. Finally, as usual, $a(t)$ is the scale factor.

The action of a universe constituted of a phantom field $\phi$,
minimally coupled to gravity, reads \cite{phant}:
\begin{equation}
S=\int d^{4}x \sqrt{-g} \left[\frac{R}{16\pi G}
+\frac{1}{2}g^{\mu\nu}\partial_{\mu}\phi\partial_{\nu}\phi+V(\phi)
+L_\text{m}\right], \label{actionquint}
\end{equation}
where $V(\phi)$ is the phantom field potential, $R$ the Ricci
scalar and $G$ the gravitational constant. The term $L_\text{m}$
accounts for the total (dark plus baryonic) matter content of the
universe, which is assumed to be a barotropic fluid with energy
density $\rho_m$ and pressure $p_m$, and equation-of-state
parameter $w_m=p_m/\rho_m$. Finally, since we focus on small
redshifts the radiation sector is neglected, although it could be
straightforwardly included.

The Friedmann equations, in units where the speed of light is 1,
write:
\begin{eqnarray}\label{FR1}
H^{2}&=&\frac{8 \pi G}{3}\Big(\rho_{m}+\rho_{\phi}\Big)-
\frac{k }{a^2}\\
\label{FR2} \dot{H}&=&-4 \pi
G\Big(\rho_{m}+p_m+\rho_{\phi}+p_{\phi}\Big)+\frac{k}{a^2},
\end{eqnarray}
where a dot denotes the derivative with respect to $t$ and
$H\equiv\dot{a}/a$ is the Hubble parameter. In these expressions,
$\rho_{\phi}$ and $p_\phi$  are respectively the energy density
and pressure of the phantom field, which are given by:
\begin{eqnarray}
\label{rhophi}
 &&\rho_{\phi} = -\frac{1}{2}\dot{\phi}^{2} + V(\phi)\\
 &&p_{\phi} = -\frac{1}{2}\dot{\phi}^{2} - V(\phi).
 \label{pphi}
\end{eqnarray}
 The evolution equation of the phantom field, describing its energy conservation
as the universe expands, is
\begin{equation}
\dot{\rho}_\phi+3H(\rho_\phi+p_\phi)=0,
\end{equation}
or written equivalently in field terms:
\begin{equation}
\ddot{\phi} + 3H\dot{\phi}- \frac{dV}{d \phi} = 0.
  \end{equation}
Note that as we mentioned in the Introduction, in phantom
cosmology the dark energy sector is attributed to the phantom
field, that is $\rho_{DE}\equiv\rho_\phi$ and $p_{DE}\equiv
p_\phi$, and thus its equation-of-state parameter is given by
\begin{eqnarray}
\label{wde}
w_{DE}\equiv\frac{p_{DE}}{\rho_{DE}}=\frac{p_\phi}{\rho_\phi}.
\end{eqnarray}
Finally, the equations close by considering the evolution of the
matter density:
\begin{eqnarray}\label{sys3}
\dot{\rho}_m+3H(1+w_m)\rho_m=0,
\end{eqnarray}
with straightforward solution
\begin{equation}
\label{energydensity}
  \rho_m = \frac{\rho_{m0}}{a^n},
  \end{equation}
where $n \equiv 3 (1 + w_m)$ and $\rho_{m0} \geq 0$ is the value
at present time $t_0$.

Lastly, note that we can extract two helpful relations, namely
from (\ref{FR1}) we obtain
\begin{equation}\label{FR1b}
\rho_{\phi}=\frac{3}{8\pi G}\Big(H^{2}-\frac{8 \pi
G}{3}\rho_{m}+\frac{k }{a^2}\Big),
\end{equation}
while from (\ref{FR2}),(\ref{rhophi}) we acquire
\begin{equation}\label{FR2b}
 \dot{\phi}^2= \frac{1}{4\pi G}\Big(\dot{H}-\frac{k
}{a^2}\Big)+\rho_m \frac{n}{3}.
\end{equation}

Let us now incorporate the power-law behavior of the scale factor.
In the case of quintessence cosmology, the power-law ansatz takes
the usual form
\begin{equation}
\label{scalefactor} a(t) = a_0 \left( \frac{t}{t_0} \right)^\beta,
\end{equation}
with $a_0$ the value of the scale factor at present time $t_0$.
However, in the case of phantom scenario, the power-law ansatz
must be slightly modified, in order to acquire self-consistency.
In particular, one rescales time as $t\rightarrow t_s-t$, with
$t_s$ a sufficiently positive reference time, and thus the scale
factor becomes \cite{Nojiri:2005vv,Nojiri:2005pu}:
\begin{equation}
\label{scalefactorphantom}
 a(t) = a_0 \left( \frac{t_s-t}{t_s-t_0}
\right)^\beta,
\end{equation}
while the Hubble parameter and its time-derivative read:
\begin{eqnarray}
\label{Ht}
 &&H(t) \equiv \frac{\dot{a}(t)}{a(t)} =- \frac{\beta}{t_s-t}\\
 && \dot{H} =
 -\frac{\beta}{(t_s-t)^2}.
 \end{eqnarray}
Therefore, for $\beta<0$ we have an accelerating ($\ddot{a}(t)>0$)
and expanding ($\dot{a}(t)>0$) universe, which possesses
additionally a positive $\dot{H}(t)$ that is it exhibits
super-acceleration \cite{Das:2005yj}. That is, in phantom
power-law cosmology, expansion is always accompanied by
acceleration. Furthermore, with $\beta<0$, at $t=t_s$ the scale
factor and the Hubble parameter diverge, that is the universe
results to a Big Rip. These behaviors are common in phantom
cosmology \cite{phant,Briscese:2006xu} and their realization is a
self-consistency test of our work. On the other hand, note that
the quintessence-ansatz (\ref{scalefactor}) cannot lead to
acceleration or to Big Rip and this was the reason for the
introduction of the phantom power-law ansatz
(\ref{scalefactorphantom}) in \cite{Nojiri:2005vv,Nojiri:2005pu}.

Having introduced the power-law ansatz that is suitable for
phantom cosmology, we can easily extract the time-dependence of
the various quantities, which re-expressed as functions of the
redshift can be confronted by the observational data. In
particular, substituting (\ref{energydensity}),(\ref{FR1b}),
(\ref{FR2b}) in (\ref{rhophi}) we obtain
\begin{equation}
\label{vraw}
 V(\phi) = \frac{3}{8 \pi G} \left( H^2 +
\frac{\dot{H}}{3} + \frac{2 k }{3a^2} \right) + \left( \frac{n -
6}{6} \right) \frac{\rho_{m0}}{a^n}.
\end{equation}
In the following we consider as usual the matter (dark plus
baryonic) component to be dust, that is $w_m\approx0$ or
equivalently $n=3$. Thus, using the ansatz
(\ref{scalefactorphantom}), and restoring the SI units using also
$M_\mathrm{P}^2 =\hbar c / 8\pi G$, we obtain
\begin{eqnarray}
 V(t) = \frac{M_\mathrm{P}^2 c}{\hbar}
 \left[ \frac{3\beta^2 - \beta}{(t_s-t)^2} + \frac{2 kc^2 (t_s-t_0)^{2\beta}}{a_0^2 (t_s-t)^{2\beta}}
 \right]
- \frac{ \rho_{m0}c^2}{2}\frac{(t_s-t_0)^{3\beta}}{a_0^3
(t_s-t)^{3\beta}}. \label{vt}
 \end{eqnarray}
Additionally, solving equation (\ref{FR2b}) for the phantom field
and inserting the power-law scale factor, gives
\begin{equation}
\label{phisolution} \phi(t) = \int \sqrt{  \frac{2 M_\mathrm{P}^2
c}{\hbar}
 \left[ -\frac{\beta}{(t_s-t)^2} - \frac{kc^2 (t_s-t_0)^{2\beta}}{a_0^2 (t_s-t)^{2\beta}} \right]
 +
  \frac{\rho_{m0}c^2 (t_s-t_0)^{3\beta}}{a_0^3 (t_s-t)^{3\beta}}}\, d
 t.
  \end{equation}
Finally, the time-dependence of the phantom energy density and
pressure can be extracted from (\ref{rhophi}) and (\ref{pphi}),
using (\ref{vt}) and (\ref{phisolution}), namely:
\begin{eqnarray}
\rho_{\phi} = \frac{M_{\rm P}^2 c}{\hbar}\left[ \frac{3
\beta^2}{(t_s-t)^2} + \frac{3 k c^2 (t_s-t_0)^{2\beta}}{a_0^2
(t_s-t)^{2\beta}} \right]  -  \frac{\rho_{m0} c^2 (t_s-t_0)^{3\beta}}{ a_0^3
(t_s-t)^{3\beta}} \label{eq_rhophi}
\end{eqnarray}
\begin{eqnarray}
 p_{\phi} = -\frac{M_{\rm P}^2
c}{\hbar}\left[ \frac{3 \beta^2 -3\beta}{(t_s-t)^2}  \right]
-\frac{ \rho_{m0} c^2 (t_s-t_0)^{3\beta}}{2 a_0^3
(t_s-t)^{3\beta}}, \label{eq_Pphi}
  \end{eqnarray}
  and thus we can straightforwardly extract the
time evolution of the dark energy equation-of-state parameter
through (\ref{wde}) as $w_{DE}(t)={p_\phi(t)}/{\rho_\phi(t)}$.
Note that at $t\rightarrow t_s$, apart from the scale factor,
$\rho_{\phi}$ and $ p_{\phi} $ diverge too, however $w_{DE}$
remains finite. This is exactly the Big Rip behavior according to
the classification of singularities of \cite{BigRip2}.

All the aforementioned time-dependencies can be expressed in terms
of the redshift $z$. In particular, since $1 + z =a_0/a$, in
phantom power-law cosmology we have
\begin{eqnarray}
 t = t_s-(t_s-t_0) (1+z)^{-1/\beta}.  \label{eq_z}
\end{eqnarray}
Therefore, using this relation we can extract the $z$-dependence
of all the relevant quantities of the scenario at hand, which can
then straightforwardly be confronted by the data.

\section{Observational constraints}
\label{Observational constraints}

In the previous section we presented the cosmological scenario in
which the dark energy sector is attributed to a phantom scalar
field, and where the scale factor is a power law of the cosmic
time. Thus, in the present section we can proceed to confrontation
with observations. In particular, we use Cosmic Microwave
Background (CMB), Baryon Acoustic Oscillations (BAO) and
Observational Hubble Data ($H_0$), in order to impose constraints
on the model parameters, and especially to the power-law exponent
$\beta$ and to the Big Rip time $t_s$. Finally, we first obtain
our results using only the CMB-WMAP7 data \cite{Larson:2010gs},
and then we perform a combined fit using additionally the BAO
\cite{Percival:2009xn} and $H_0$ ones \cite{ref:0905}.

We mention that in the present work we prefer not to use SNIa data
as in the combined WMAP5+BAO+SNIa dataset \cite{Komatsu:2008hk}.
This is because the combined WMAP5 dataset uses SNIa data from
\cite{Hicken:2009dk, Kowalski:2008ez} which do not include
systematic error, and the cosmological parameters derived from
the combined WMAP5 dataset also differ from those derived from
other compilations of SNIa data \cite{Kessler:2009ys}.
Inclusion of the SNIa systematic error which is comparable to the
its statistical error can significantly alter the value of the
equation of state \cite{Komatsu:2010fb}. Furthermore, recent
analysis shows that the value of the equation-of-state parameter
derived from two different light-curve fitters could be different
from the one derived from two different datasets. This could make
it difficult to identify if $w_{DE}$ is phantom, since its
obtained values from the two fitters are different
\cite{Bengochea:2010it}. A very recent critics on SNIa data
analysis has been presented in \cite{Vishwakarma:2010nc}.
Definitely the incorporation of SNIa data in constraining phantom
cosmology is a subject that deserves further investigation.

Similarly to the non-phantom case \cite{Thepsuriya:2009wq}, the
exponent $\beta$ can be straightforwardly expressed as
 \begin{eqnarray}
  \beta = -H_0(t_s-t_0),
   \end{eqnarray}
where, as usual, we use the subscript 0 to denote the value of a
quantity at present, and we moreover set $a_0$ to 1. Furthermore,
we introduce the usual density parameter $\Omega_m\equiv8\pi G
\rho_m/(3H^2)$, and we split $\Omega_m$ in its baryonic and cold
dark matter part, $\Omega_b$ and $\Omega_{CDM}$ respectively
($\Omega_m=\Omega_b+\Omega_{CDM}$). Lastly, it proves convenient
to introduce the critical density $\rho_{c}=3 H^2/8 \pi G$, and
thus we can use the relation $ \rho_{m0} = \Omega_{m0} \rho_{c0}$.

In a general, non-flat geometry the Big Rip time $t_s$ cannot be
calculated, bringing a large uncertainty to the observational
fitting. However, one could estimate it, performing some very
plausible assumptions \cite{Caldwell:2003vq}. In particular,
assuming a flat geometry, which is a very good approximation
\cite{Komatsu:2010fb}, and assuming that at late times the phantom
dark energy will dominate the universe, which is always the case
in phantom models, $t_s$ can be expressed as
\cite{Caldwell:2003vq}
 \begin{eqnarray}
 t_s \simeq
t_0+\frac{2}{3}|1+w_{DE}|^{-1}H_0^{-1}(1-\Omega_{m0})^{-1/2}.
\label{tsestimation}
\end{eqnarray}
Here we have to mention that there is one last assumption in
extracting this relation, namely that at late times the dark
energy equation-of-state parameter $w_{DE}$ approaches a constant
value. Fortunately, this is always the case in flat power-law
phantom cosmology examined in this work, as can be seen from
(\ref{eq_rhophi}), (\ref{eq_Pphi}) for $k=0$, recalling also that
$\beta$ is always negative in an expanding universe. In this case,
at late times we indeed have:
 \begin{eqnarray}
w_{DE}\simeq-1+\frac{1}{\beta}, \label{wDEestimation}
\end{eqnarray}
which lies always below the phantom divide as
expected\footnote{Note that if instead of $w_{DE}$ we consider the
effective $w_{\rm eff}$, that is including the weighted
contribution of matter, then we have
 \begin{eqnarray}
w_{\rm eff}\rightarrow-1+\frac{2}{3\beta} \label{weffestimation}
\end{eqnarray}
at $t\rightarrow t_s$ \cite{Gumjudpai:2008mg}, for any curvature
value.}. In addition, one can straightway extract $[H(t)^2]$
through (\ref{Ht}) as
 \begin{eqnarray}
[H(t)^2]=H_0^2\left(\frac{t_s-t_0}{t_s-t}\right)^2.
\label{Htsquare}
\end{eqnarray}
Finally, as we have mentioned, the time-functions can be expressed
as redshift-functions using (\ref{eq_z}).

Having all the required information, we proceed to the data
fitting. For the case of the WMAP7 data alone we use the maximum
likelihood parameter values for $H_0$, $t_0$,
$\Omega_{\mathrm{CDM}0}$ and $\Omega_{b0}$ \cite{Komatsu:2010fb},
focusing on the flat geometry. Additionally, we perform a combined
observational fitting, using WMAP7 data, along with Baryon
Acoustic Oscillations (BAO) in the distribution of galaxies, and
Observational Hubble Data ($H_0$). The details and the techniques
of the construction are presented in the Appendix.

\section{Results and Discussions}
\label{results}

In the previous section we presented the method that allows for
the confrontation of power-law phantom cosmology with the data. In
the present section we perform such an observational fitting,
presenting our results, and discussing their physical
implications.

First of all, in Table \ref{datatable}, we show for completeness
the maximum likelihood values for the present time $t_0$, the
present Hubble parameter $H_0$, the present baryon density
parameter $\Omega_{b0}$ and the present cold dark matter density
parameter $\Omega_{\mathrm{CDM}0}$, that was used in our fitting
\cite{Komatsu:2010fb}, in WMAP7 as well as in the combined
fitting.
\begin{table*}[!]
\begin{ruledtabular}
\begin{tabular}{ccc}
\textbf{Parameter} & \textbf{WMAP7+BAO+$H_0$} & \textbf{WMAP7}\\
\hline\hline
$t_0$ & $13.78\pm0.11$ Gyr [$(4.33\pm0.04) \times 10^{17}$ sec] & $13.71\pm0.13$  Gyr  [$(4.32\pm0.04) \times 10^{17}$ sec] \\
$H_0$ & $70.2^{+1.3}_{-1.4}$ km/s/Mpc & $71.4\pm2.5$ km/s/Mpc\\
$\Omega_{b0}$ & $0.0455\pm0.0016$ & $0.0445\pm0.0028$\\
$\Omega_{\mathrm{CDM}0}$ & $0.227\pm0.014$ & $0.217\pm0.026$\\
\end{tabular}
\caption{Observational maximum likelihood values in 1$\sigma$
confidence level for the present time $t_0$, the present Hubble
parameter $H_0$, the present baryon density parameter
$\Omega_{b0}$ and the present cold dark matter density parameter
$\Omega_{\mathrm{CDM}0}$, for WMAP7 as well as for the combined
fitting WMAP7+BAO+$H_0$. The values are taken from
\cite{Komatsu:2010fb}.} \label{datatable}
\end{ruledtabular}
\end{table*}
In the same Table we also provide the 1$\sigma$ bounds of every
parameter. In Table \ref{resulttable} we present the maximum
likelihood values and the 1$\sigma$ bounds for the derived
parameters, namely the power-law exponent $\beta$, the present
matter energy density value $\rho_{m0}$, the present critical
energy density value $\rho_{c0}$ and the Big Rip time $t_s$.
\begin{table*}[!]
\begin{ruledtabular}
\begin{tabular}{ccc}
    \textbf{Parameter} & \textbf{WMAP7+BAO+$H_0$} & \textbf{WMAP7}\\
    \hline \hline
    $\beta$ & $-6.51^{+0.24}_{-0.25}$ & $-6.5\pm0.4$\\
    $\rho_{m0}$ & $(2.52\pm0.26) \times 10^{-27}$ ${\rm kg/m^3}$ & $(2.50\pm0.30) \times 10^{-27}$ ${\rm kg/m^3}$\\
    $\rho_{c0}$ & $(9.3^{+0.3}_{-0.4})\times 10^{-27}$ ${\rm kg/m^3}$ & $(9.57\pm0.67) \times 10^{-27}$ $ {\rm
    kg/m^3}$\\
        $t_s$ & $104.5^{+1.9}_{-2.0}$ Gyr [$(3.30\pm0.06) \times 10^{18}$ sec] & $102.3\pm3.5$ Gyr [$(3.23\pm0.11) \times 10^{18}$ sec] \\
\end{tabular}
\caption{Derived maximum likelihood values in 1$\sigma$ confidence
level for the power-law exponent $\beta$, the present matter
energy density value $\rho_{m0}$, the present critical energy
density value $\rho_{c0}$ and the Big Rip time $t_s$, for WMAP7 as
well as for the combined fitting WMAP7+BAO+$H_0$.}
\label{resulttable}
\end{ruledtabular}
\end{table*}
As we observe, $\beta$ is negative, as expected in consistent
phantom cosmology. We mention here that the phantom power-law
ansatz (\ref{scalefactorphantom}) is technically different from
the quintessence one (\ref{scalefactor}), and thus one cannot
straightforwardly compare the exponent values of the two cases
(for example a similar $w_{DE}$ is produced by significantly
different exponents in the two scenarios \cite{Sethi2005}). Now,
note that the Big Rip time is one order of magnitude larger than
the present age of the universe, which shows that such an outcome
is unavoidable in phantom cosmology, unless one include additional
mechanisms for the exit from phantom phase \cite{Sami:2003xv}, an
approach that was not taken into account in this work.

Let us discuss in more detail the values and the evolution of some
quantities of interest. For the combined data WMAP7+BAO+$H_0$, the
potential (\ref{vt}) is fitted as
\begin{eqnarray}
    V(t) \approx \frac{6.47\times 10^{27}}{(3.30\times
    10^{18}-t)^2}  -  2.51\times 10^{-371}(3.30\times
    10^{18}-t)^{19.54},
     \label{Vtcomb}
   \end{eqnarray}
while WMAP7 data alone give
   \begin{eqnarray}
      \label{Vtalone}
    V(t)\approx\frac{6.37\times 10^{27}}{(3.23\times
    10^{18}-t)^2}   -  1.99\times 10^{-368}(3.23\times 10^{18}-t)^{19.39}.
    \end{eqnarray}
Note that the second terms in these expressions, although very
small at early times, they become significant at late times, that
is close to the Big Rip. In particular, the inflection happens at
$22.4^{+1.9}_{-2.0}$ Gyr     (WMAP7+BAO+$H_0$) and
$22.0^{+3.5}_{-3.5}$ Gyr (WMAP7), after which we obtain a rapid
increase.

Now, concerning the scalar field evolution $\phi(t)$, at late
times ($t \rightarrow t_s$) the $\rho_{m0}$-term in
(\ref{phisolution}) can be neglected. Thus, (\ref{phisolution})
reduces to
 \be
 \phi(t) \approx  \int \sqrt{ -
\frac{2 M_\mathrm{P}^2 c}{\hbar} \frac{\beta}{(t_s-t)^2}}\; d t,
 \ee  which can be fitted using combibed WMAP7+BAO+$H_0$ giving
\be
 \label{phitcomb} \phi(t)\approx -2.64\times10^{13} \ln{(3.30\times10^{18}-t)},
 \ee
while for WMAP7 dataset alone we obtain \be
\phi(t)\approx-2.63\times10^{13} \ln{(3.23\times10^{18}-t)}.
 \label{phitalone}
 \ee
 As expected, both the phantom field and its kinetic energy
 ($-\dot{\phi}^2/2$) diverge at the Big Rip.

\begin{figure}[ht]
    \begin{center}
    \includegraphics[width=9cm, height=7cm, angle=0]{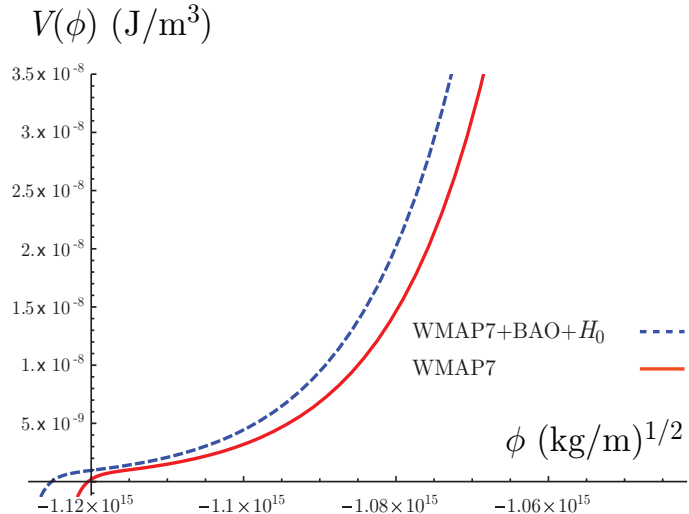}
    \end{center}
    \caption{{\it{The phantom potential obtained from observational data fitting of WMAP7 and WMAP7+BAO+$H_0$.}} }\label{potentialplots}
    \end{figure}

Having fitted the phantom potential $V(t)$ and the phantom field
itself $\phi(t)$, it is now straightforward to obtain the
potential as a function of the phantom field, namely $V(\phi)$. In
particular, (\ref{phitcomb}) and (\ref{phitalone}) can be easily
inverted, giving $t(\phi)$, and thus substitution to
(\ref{Vtcomb}) and (\ref{Vtalone}) respectively provides
$V(\phi)$. Doing so, for the combined data WMAP7+BAO+$H_0$ the
potential is fitted as
\begin{eqnarray}
    V(\phi) \approx  6.47\times 10^{27}\, e^{0.75\times 10^{-13}\phi}   -  2.51\times 10^{-371}\, e^{-7.4\times 10^{-13}\phi},
     \label{Vphicomb}
   \end{eqnarray}
while for WMAP7 dataset alone we obtain
\begin{eqnarray}
   V(\phi) \approx  6.37\times 10^{27}\, e^{0.76\times 10^{-13}\phi}   -  1.99\times 10^{-368}\, e^{-7.4\times
    10^{-13}\phi}.
     \label{Vphicomb}
   \end{eqnarray}
  In order to provide a more transparent picture, in
  Fig.~\ref{potentialplots} we present the corresponding plot for $V(\phi)$, for both the
   WMAP7+BAO+$H_0$ as well as the WMAP7 case.

Let us now consider the equation-of-state parameter for the
phantom field, that is for the dark energy sector. As we mentioned
in the end of section \ref{model}, it is given by
$w_{DE}(t)={p_\phi(t)}/{\rho_\phi(t)}$, with $p_\phi(t)$ and
$\rho_\phi(t)$ given by relations (\ref{eq_Pphi}) and
(\ref{eq_rhophi}) respectively. Finally, one can extract the
redshift dependence using (\ref{eq_z}). One can therefore use
WMAP7 and WMAP7+BAO+$H_0$ observational data in order to fit the
evolution of $w_{DE}(z)$ at late times, that is for $t\rightarrow
t_s$, or equivalently for $z\rightarrow -1$.
 For the WMAP7+BAO+$H_0$ combined dataset we find
\begin{eqnarray}
\label{wzcomb}
 w_{DE}(z)&\approx&\frac{1}{2}-\frac{6.068}{3.670- (1+z)^{3.307}},
\end{eqnarray}
while for the WMAP7 dataset alone we have
\begin{eqnarray}
\label{wzalone}
 w_{DE}(z)&\approx&\frac{1}{2}-\frac{6.328}{3.824- (1+z)^{3.309}}.
\end{eqnarray}
As we observe, at $t\rightarrow t_s$, $w_{DE}$ becomes -1.153 for
the combined dataset and -1.155 for the WMAP7 dataset alone.
However, as we have already discussed in the end of section
\ref{model}, at $t\rightarrow t_s$ despite the finiteness of $
w_{DE}$, the phantom dark energy density and pressure become
infinite. These behaviors are the definition of a Big Rip
\cite{BigRip2}, and this acts as a self-consistency test of our
model.

\section{Conclusions}
\label{Conclusions}

In this work we investigated phantom cosmology in which the scale
factor is a power law. After constructing the scenario, we used
observational data in order to impose constraints on the model
parameters, focusing on the power-law exponent $\beta$ and on the
Big Rip time $t_s$.

Using the WMAP7 dataset alone, we found that the power-law
exponent is $\beta\approx-6.5\pm0.4$  while the Big Rip is
realized at $t_s\approx 102.3\pm3.5$ Gyr, in 1$\sigma$ confidence
level. Additionally, the dark-energy equation-of-state parameter
$w_{DE}$ lies always below the phantom divide as expected, and at
the Big Rip it remains finite and equal to -1.155. However, both
the phantom dark-energy density and pressure diverge at the Big
Rip.

Using WMAP7+BAO+$H_0$ combined observational data we found that
$\beta\approx-6.51^{+0.24}_{-0.25}$, while
$t_s\approx104.5^{+1.9}_{-2.0}$ Gyr,  in 1$\sigma$ confidence
level. Moreover, $w_{DE}$ at the Big Rip becomes -1.153. Finally,
in order to present a more transparent picture, we provided the
reconstructed phantom potential.

In summary, we observe that phantom power-law cosmology can be
compatible with observations, exhibiting additionally the usual
phantom features, such is the future Big Rip singularity. However,
it exhibits also the known disadvantage that the dark-energy
equation-of-state parameter lies always below the phantom divide,
by construction. In order to acquire a more realistic picture,
describing also the phantom divide crossing, as it might be the
case according to observations, one should proceed to the
investigation of quintom power-law cosmology, considering apart
from the phantom a canonical scalar field, too. Such a project is
left for future investigation.

Finally, let us make a comment on the nature of the investigated
scenarios. Although the classical behavior of phantom fields has a very
rich phenomenology and can be compatible with observations, as it is known
the discussion about the
construction of quantum field theory of phantoms is still open in
the literature. For instance in \cite{Cline:2003gs} the authors
reveal the causality and stability problems and the possible
spontaneous breakdown of the vacuum into phantoms and conventional
particles in four dimensions. However, on the other hand, there have also
been serious attempts in overcoming these difficulties and construct a
phantom theory consistent with the basic requirements of quantum field
theory \cite{quantumphantom0}, with the phantom fields arising as an
effective description. The present analysis is just a first approach on
phantom power-law cosmology. Definitely, the subject of quantization of
such scenarios is open and needs further investigation.

\section*{Acknowledgments}
We thank Kiattisak Thepsuriya and the referee for useful discussions and comments.  C.~K. is
supported by a research studentship funded by Thailand Toray
Science Foundation (TTSF) and the Thailand Center of Excellence in Physics (ThEP). B.~G. is sponsored by the Thailand
Research Fund's Basic Research Grant (TRF Advanced Research
Scholar), TTSF and ThEP.

\appendix*

\section{Observational data and constraints}
\label{Observational data and constraints}

In this Appendix we briefly review the main sources of
observational constraints used in this work, namely WMAP7 Cosmic
Microwave Background (CMB), Baryon Acoustic Oscillations (BAO),
and Observational Hubble Data ($H_0$). In our calculations we take
the total likelihood $L\propto e^{-\chi^2/2}$ to be the product of
the separate likelihoods of BAO, CMB and $H_0$. Thus, the total
$\chi^2$ is
\begin{eqnarray}
\chi^2(p_s)=\chi^2_{CMB}+\chi^2_{BAO}+\chi^2_{H_0}.
\end{eqnarray}
\\

{\it{a. CMB constraints}}\\

We use the CMB data to impose constraints on the parameter space,
following the recipe described in \cite{Komatsu:2008hk}.
 The ``CMB
shift parameters'' \cite{Wang1} are defined as:
 \be
  R\equiv
\sqrt{\Omega_{m0}}H_0 r\(z_*\),\,\quad l_{a}\equiv \pi
r\(z_*\)/r_{s}\(z_*\).
 \ee
  $R$ can be physically interpreted as a
scaled distance to recombination, and $l_{a}$ can be interpreted
as the angular scale of the sound horizon at recombination. $r(z)$
is the comoving distance to redshift $z$ defined as
 \be
r(z)\equiv\int_{0}^{z}\frac{1}{H\(z\)}dz,
 \ee
 while $r_{s}\(z_*\)$ is
the comoving sound horizon at decoupling (redshift $z_*$), given
by
 \be
r_{s}\(z_*\)=\int_{z_*}^{\infty}\frac{1}{H\(z\)\sqrt{3\(1+R_{b}/\(1+z\)
\)}}dz.
 \ee
  The quantity $R_b$ is the ratio of the energy density
of photons to baryons, and its value can be calculated as
$R_b=31500 \Omega_{b0} h^2 \(T_{CMB}/2.7K\)^{-4}$, ($\Omega_{b0}$
being the present day density parameter for baryons) using
$T_{CMB}=2.725$  \cite{Komatsu:2008hk}. The redshift at decoupling
$z_*\(\Omega_{b0},\Omega_{m0},h\)$ can be calculated from the
following fitting formula \cite{husugiyama}:
 \be
z_*=1048\[1+0.00124\(\Omega_{b0}
h^2\)^{-0.738}\]\[1+g_1\(\Omega_{m0} h^2\)^{g_2}\], \ee with $g_1$
and $g_2$ given by:
\begin{eqnarray*}
g_1&=&\frac{0.0783\(\Omega_{b0} h^2\)^{-0.238}}{1+39.5\(\Omega_{b0} h^2\)^{0.763}}\\
g_2&=&\frac{0.560}{1+21.1\(\Omega_{b0} h^2\)^{1.81}}.
\end{eqnarray*}
Finally, the $\chi^2$ contribution of the CMB reads
 \be
\chi^{2}_{CMB}=\mathbf{V}_{\rm CMB}^{\mathbf{T}}\mathbf{C}_{\rm
inv}\mathbf{V}_{\rm CMB}. \ee
 Here $\mathbf{V}_{\rm
CMB}\equiv\mathbf{P}-\mathbf{P}_{\rm data}$, where $\mathbf{P}$ is
the vector $\(l_{a},R,z_{*}\)$ and the vector $\mathbf{P}_{\rm
data}$ is formed from the WMAP $5$-year maximum likelihood values
of these quantities \cite{Komatsu:2008hk}. The inverse covariance
matrix $\mathbf{C}_{\rm inv}$ is also provided in
\cite{Komatsu:2008hk}.
\\

{\it{b. Baryon Acoustic Oscillations constraints}}\\

In this case the measured quantity  is the ratio
$d_z=r_{s}\(z_{d}\)/D_{V}\(z\)$, where $D_{V}\(z\)$ is the so
called ``volume distance'', defined in terms of the angular
diameter distance $D_{A}\equiv r\(z\) /\(1+z\)$ as \be
D_{v}\(z\)\equiv\left[\frac{\(1+z\)^2 D_{A}^{2}(z) z
}{H(z)}\right]^{1/3},
 \ee
and $z_d$ is the redshift of the baryon drag epoch, which can be
calculated from the fitting formula \cite{HuEisenstein}: \be
z_d=\frac{1291\(\Omega_{m0} h^2\)^{0.251}}{1+\(\Omega_{M0}
h^2\)^{0.828}}\[1+b_1\(\Omega_{b0} h^2\)^{b_2}\],
 \ee
where $b_1$ and $b_2$ are given by
\begin{eqnarray*}
b_1&=&0.313\(\Omega_{m0} h^2\)^{-0.419}\[1+0.607\(\Omega_{m0} h^2\)^{0.674}\]\\
b_2&=&0.238\(\Omega_{m0} h^2\)^{0.223}.
\end{eqnarray*}

We use the two measurements of $d_z$ at redshifts $z=0.2$ and
$z=0.35$ \cite{Percival:2009xn}. We calculate the $\chi^2$
contribution of the BAO measurements as: \be
\chi^{2}_{BAO}=\mathbf{V}_{\rm BAO}^{\mathbf{T}}\mathbf{C}_{\rm
inv}\mathbf{V}_{\rm BAO}.
 \ee
    Here the vector $\mathbf{V}_{\rm
BAO}\equiv\mathbf{P}-\mathbf{P}_{\rm data}$, with
$\mathbf{P}\equiv \( d_{0.2},d_{0.35} \) $, and $\mathbf{P}_{\rm
data}\equiv\(0.1905, 0.1097\)$, the two measured BAO data points
\cite{Percival:2009xn}. The inverse covariance matrix is provided
in \cite{Percival:2009xn}.\\

{\it{c. Observational Hubble Data constraints}}\\

The observational Hubble data are based on differential ages of
the galaxies \cite{ref:JL2002}. In \cite{ref:JVS2003}, Jimenez
{\it et al.} obtained an independent estimate for the Hubble
parameter using the method developed in \cite{ref:JL2002}, and
used it to constrain the equation of state of dark energy. The
Hubble parameter, depending on the differential ages as a function
of the redshift $z$, can be written as
\begin{equation}
H(z)=-\frac{1}{1+z}\frac{dz}{dt}.
\end{equation}
Therefore, once $dz/dt$ is known, $H(z)$ is directly obtained
\cite{ref:SVJ2005}. By using the differential ages of
passively-evolving galaxies from the Gemini Deep Deep Survey
(GDDS) \cite{ref:GDDS} and archival data \cite{ref:archive1},
Simon {\it et al.} obtained $H(z)$ in the range of $0\lesssim z
\lesssim 1.8$ \cite{ref:SVJ2005}. We use the twelve observational
Hubble data from \cite{ref:0905}  listed in Table
\ref{Hubbledata}.
\begin{table}[htbp]
\begin{center}
\begin{tabular}{c|llllllllllll}
\hline\hline
 $z$ &\ 0 & 0.1 & 0.17 & 0.27 & 0.4 & 0.48 & 0.88 & 0.9 & 1.30 & 1.43 & 1.53 & 1.75  \\ \hline
 $H(z)\ ({\rm km~s^{-1}\,Mpc^{-1})}$ &\ 74.2 & 69 & 83 & 77 & 95 & 97 & 90 & 117 & 168 & 177 & 140 & 202  \\ \hline
 $1 \sigma$ uncertainty &\ $\pm 3.6$ & $\pm 12$ & $\pm 8$ & $\pm 14$ & $\pm 17$ & $\pm 60$ & $\pm 40$
 & $\pm 23$ & $\pm 17$ & $\pm 18$ & $\pm 14$ & $\pm 40$ \\
\hline
\end{tabular}
\end{center}
\caption{\label{Hubbledata} The observational $H(z)$
data~\cite{ref:0905}.}
\end{table}

 The best-fit values of the model parameters from
observational Hubble data \cite{ref:SVJ2005} are determined by
minimizing
\begin{equation}
\chi_{H_0}^2(p_s)=\sum_{i=1}^{12} \frac{[H_{th}(p_s;z_i)-H_{
obs}(z_i)]^2}{\sigma^2(z_i)},\label{eq:chi2H}
\end{equation}
where $p_s$ denotes the parameters contained in the model,
$H_{th}$ is the predicted value for the Hubble parameter,
$H_{obs}$ is the observed value, $\sigma(z_i)$ is the standard
deviation measurement uncertainty, and the summation runs over the
$12$ observational Hubble data points at redshifts $z_i$.

\end{document}